\documentstyle[twocolumn,prb,aps,psfig]{revtex}

\begin{document}

\title{Hierarchy of Fast Motions in Protein Dynamics}
\author{Alexey K. Mazur}
\address{Laboratoire de Biochimie Th\'eorique, CNRS UPR9080\\
Institue de Biologie Physico-Chimique\\
13, rue Pierre et Marie Curie, Paris,75005, France.\\
FAX:(33-1) 43.29.56.45. Email: alexey@ibpc.fr}

\address{\medskip\em
\begin{minipage}{14cm}
{}~~~
For many biological applications of molecular dynamics (MD) the importance of
good sampling in conformational space makes it necessary to eliminate the
fastest motions from the system in order to increase the time step. An
accurate knowledge of these motions is a necessary prerequisite for such
efforts. It is known that harmonic vibrations of bond lengths and bond angles
produce the highest frequencies in proteins. There are also fast anharmonic
motions, such as inter-atomic collisions, which are probably most important
when bond lengths and bond angles are fixed. However, the specific time
scales corresponding to all these limitations are not known precisely. In
order to clarify the above issue this paper analyses time step limiting
factors in a series of numerical tests by using an internal coordinate
molecular dynamics approach, which allows chosen internal coordinates to be
frozen. It is found that, in proteins, there is a rather complicated
hierarchy of fast motions, with both harmonic and anharmonic effects mixed
together at several close time scales. Non-bonded interactions, notably
strong hydrogen bonding, create locally distributed normal modes with
frequencies similar to those of bond stretching between non-hydrogen atoms.
They also impose ubiquitous anharmonic limitations starting
from rather small step sizes. With fixed standard amino acid geometry,
rotations of hydrogen bonded hydroxyl groups limit time steps at the 5 fsec
level. The next important limitation occurs around 10 fsec and is created by
collisions between non-hydrogen atoms.
\end{minipage} }
%\date{\today}
\maketitle

\section{Introduction}
In the literature devoted to protein dynamics two classes of motions
are most frequently discussed. The first comprises the slowest motions
involved in
the folding and various biological functions of proteins. The second class is at
the opposite border of the spectrum and includes the fastest motions which
limit time steps in molecular dynamics (MD) calculations and, consequently,
the time scales accessible for simulation. Because of the high computer cost
of long MD trajectories for biological systems, considerable efforts have
always been applied to overcome these limitations.
The first and most popular method is constraint
MD\cite{SHAKE:,vanGunsteren:77,Ciccotti:86}. Among the techniques
developed in recent years one can mention, for example,
multiple time scale MD\cite{Pinches:78,Tuckerman:92}, weighted mass
MD\cite{Pomes:90,Mao:91} and internal coordinate MD (ICMD)
\cite{Pear:79,BKS0:,Gibson:90,Jain:93,Rice:94}. Within the framework of this
research the physical nature and the specific time scales of the fast
motions in proteins have often been discussed \cite{Brooks:88}.
There are several reasons, however, why this issue required the more
complete analysis presented here.

One of the motivations came from recent progress in ICMD, that is
dynamics simulations in the space of internal rather than Cartesian
coordinates\cite{Mathiowetz:94,Mzjcc:97}. By this method, proteins can be
modeled as chains of rigid bodies with only torsional degrees of freedom. It
is not obvious {\em a priori} what are the time step limiting motions in such
models, and, in the literature, there are rather contradictory estimates of
the overall prospects of ICMD compared to constraint MD
\cite{vanGunsteren:82,Dor-jcp:93}. Another reason why this question
attracts attention came from the recent results concerning the properties of
the numerical algorithms commonly employed in MD, namely integrators of the
St{\"o}rmer-Verlet-leapfrog group\cite{Mzjcp:97}. It has been found that the
fluctuation of the computed instantaneous total energy of a microcanonical
ensemble, which is generally used as the most important test
for quality of MD trajectories, is dominated by simple interpolation errors
and cannot serve as a measure of accuracy. It is possible, therefore, that
previous estimates of limiting step sizes were significantly biased.

Proteins have hierarchical spatial organization and this structural hierarchy
is naturally mapped onto the spectrum of their motions. Namely, fast
motions involve individual atoms and chemical groups while the slow ones
correspond to displacements of secondary structures, domains etc. Every such
movement considered separately can be characterized by a certain maximum time
step, and in this sense one can say that there exists a hierarchy of fast
motions and, accordingly, of step size limits. The lowest such limit is
determined by
bond stretching vibrations of hydrogens, but for our purposes the following
few levels of this hierarchy are most interesting. Normally it is assumed
that stretching of bonds between non-hydrogen atoms forms the next
level\cite{vanGunsteren:90}. However, this intuitive suggestion is not easy
to verify because protein normal modes in this frequency range are not always
local, and their high frequencies cannot be readily attributed to
specific harmonic terms in the force field. On the other hand, normal modes
with bond lengths constrained have never been studied. It should also be
expected that, because of the limitations imposed by fast anharmonic motions,
the hierarchy does not exactly correspond to the spectra of
normal modes. All these issues are considered in detail below.

The paper is organized as a simple sequence of numerical tests on a model
protein with the fastest motions suppressed one after another.
Step size limits are
determined by a unique test proposed and analyzed in detail elsewhere
\cite{Mzjcp:97}. It is found that the hierarchy
of step size limits is rather complicated and does not always follow
common intuitive assumptions. For instance, bond stretching between
non-hydrogen atoms in fact overlaps the frequency range of certain
collective vibrations and, therefore, does not create a separate limitation
on time steps.
On the other hand, with step sizes beyond 10 fsec, anharmonic effects become
dominant. In agreement with recent studies \cite{Mzjcp:97}, leapfrog
trajectories appear to hold to correct constant-energy hypersurfaces with
considerably larger time steps than commonly recommended.

\section{Results and Discussion}
\subsection*{Test Systems and Simulation Protocols}
The main model system consists of an immunoglobulin binding domain of
streptococcal protein G \cite{pgb:} which is a 56 residue $\alpha/\beta$
protein subunit (file 1pgb in the Protein Database \cite{PDB:}) with 24 bound
water molecules available in the crystal structure. All hydrogens are
considered explicitly giving a total of 927 atoms. In order to separate effects
produced by water from those due to the protein itself similar tests were
also performed for a droplet of TIP3P water molecules. The initial
configuration of the droplet was obtained by taking the coordinates of the first
100 molecules closest to the center of the standard water box supplied with
the AMBER package \cite {AMBER:,TIP3P:}. Equilibration for low temperature
tests was performed starting from the local energy minimum corresponding to
this configuration rather than from an ice crystal structure.

In all calculations the AMBER94 force field was employed \cite{AMBER94:} without
truncation of non-bonded interactions. Molecular motions were generally
simulated with internal coordinates as independent variables by using
Hamiltonian equations of motion and an implicit leapfrog integrator described
elsewhere \cite{Mzjcc:97}. In one case, however, namely for normal
temperature calculations with fixed bond lengths to hydrogen atoms, the
standard AMBER package was employed, because
the ICMD algorithm needs too many iterations for convergence
with large step sizes \cite{Mzjcc:97}.
Comparisons between ICMD trajectories and usual Cartesian
MD show no essential differences \cite{Mzjcc:97}, therefore, for consistency
and convenience, internal coordinates have been preferred wherever possible.

Initial data for all numerical tests were prepared with the standard
protocol described earlier \cite{Mzjcc:97,Mzjcp:97} which makes possible
smooth initialization of leapfrog trajectories always from a single
constant energy hypersurface. In order to simulate a virtually harmonic
behavior parallel tests were performed at the very low
temperature of 1K for which the equilibration protocol was modified as
follows. During the first 5~psec all velocities were reassigned several times
by sampling from a Maxwell distribution with T=1~K. During the following
7.5~psec velocities were rescaled periodically if the average temperature
went above 2~K. This modification was necessary since in the virtually
harmonic low temperature conditions energy equipartition is reached
very slowly and the initial distribution over normal modes can persist for
a long time. The necessary harmonic frequencies were obtained from
calculated spectral densities of autocorrelation functions of appropriate
generalized velocities. In all production runs the duration of the test
trajectory was 10 psec.

The estimates of maximal time steps are normally made by checking
conservation of the total energy of a microcanonical ensemble
\cite{Allen:87,Haile:92}. Our approach is similar, but it accurately takes
into account certain non-trivial properties of the leapfrog discretization of
MD trajectories\cite{Mzjcp:97}. The test trajectory is repeatedly
calculated always starting from the same constant-energy hypersurface. In
each run certain system averages are evaluated and compared with ``ideal''
values, i.e. the same parameters obtained with a very small time step. The
choice of such parameters must correspond to the leapfrog discretization,
which means that they should be computed from on-step coordinates and
half-step velocities without additional interpolations. The latter condition,
together with the ``smooth start", distinguishes this approach from earlier
testing strategies. These modifications are essential because they remove a
significant and systematic bias present in the traditional approach which
employs the time fluctuation of the instantaneous total energy as an
indicator of accuracy \cite{Mzjcp:97}.

The parameters we use are as follows: the average potential energy, $\bar U$,
and its time variance, $D[U]$; the average kinetic energy, $\bar K$, computed
for half-steps; the total energy, $E=\bar U+\bar K$, and its drift computed
in the same way, referred to below as $E$-drift. For a sufficiently
long trajectory of a Hamiltonian system $\bar U$, $E$ and $D[U]$ characterize
the sampled hypersurface in phase space. Their deviations from the
corresponding virtually ideal values characterize the bias of the sampling
obtained and, therefore, can be used for accessing step size limits. The
$E$-drift computed in this way is exactly zero for ideal harmonic systems
\cite{Mzjcp:97}, and is thus a good indicator of anharmonic effects.

\begin{figure}
\centerline{\psfig{figure=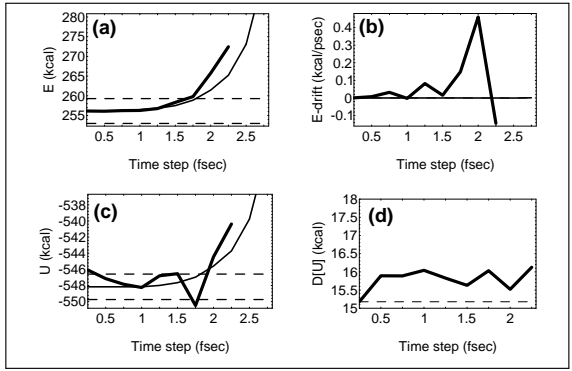,height=7.5cm,angle=0.}}
\caption{Characteristic time step dependencies obtained for a completely free
protein. The corresponding low temperature plots are shown by thinner lines
in (a-c). For comparison, in (a) and (c) the low temperature deviations for
the total and potential energies had been scaled and shifted to fit the range
of deviations observed with normal temperature. The dashed horizontal lines
in (a) and (c) show the bands of acceptable deviation corresponding to
variances $D\lbrack U\rbrack$ indicated by the dashed line in (d). The
definition of these bands is given in the text.}\label{Ffree}\end{figure}

As an example let us consider results of such testing for a completely free
protein.  It is seen in Fig. \ref{Ffree} that, below a step size
of about 1.7~fsec, all the measured parameters remain approximately constant
and close to the accurate values. Above this level some
deviations grow rapidly. This characteristic behavior is similar to that of a
simple leapfrog harmonic oscillator \cite{Mzjcp:97,Hockney:81}. All its
properties depend upon the reduced step size $\tau =\omega h$, where $\omega$
is the frequency, and it can be shown analytically that, with small $\tau$,
power series expansions for deviations of $E$ and $\bar U$ are
dominated by the terms
of the fourth and sixth orders \cite{Mzjcp:97}. This explains
why the deviations grow rapidly beyond a certain threshold which depends
mainly upon the highest frequencies in the system.

In order to access the accuracy quantitatively we need to compare the
deviations of average energies with some scale. An appropriate natural scale
is given by the time variations of the potential energy characterized by the
value of $D[U]$ shown in Fig. \ref{Ffree}~(d). We will take a deviation of
$0.1D[U]$ as the upper acceptable level for $\bar U$ and a two times larger
value for $E$. In a harmonic system, $\bar U$ and $\bar K$ computed as
described are equal \cite{Mzjcp:97}; so the deviation of $E$ is exactly two
times that of $\bar U$ and it reaches its upper level simultaneously with
$\bar U$, i.e. with the
same characteristic step size denoted as $h_{c}$. The threshold levels chosen
are certainly arbitrary, but they are reasonable and normally $h_{c}$ values
appear similar to maximal step sizes reported in the literature. We note,
however, that we will be mainly interested here in relative rather than in
absolute $h_{c}$ values for different models.

Thus, the dotted line in Fig. \ref{Ffree}~(d) marks the best estimate of the
variance $D[U]$ and similar lines in Figs. \ref{Ffree}~(a,c) show the
corresponding acceptance intervals for energies. It can be seen in
Figs. \ref{Ffree} (a) and (c) that, at normal temperature, the deviations
of the total and potential energies look qualitatively different because
$\bar U$ is affected by occasional transitions between local minima which are
stochastic and cannot be properly averaged during the relatively short test
trajectory. Nevertheless, both $\bar U$ and $E$ go beyond their acceptance
intervals with $h_{c}\approx1.7$~fsec. At low temperature, both deviations are
regular, and they yield almost exactly the same $h_c$ value. The highest
frequencies in this system are those of the bond stretching modes of hydrogens
and they range from approximately 3000 cm$^{-1}$ for aliphatic groups
to 3800 cm$^{-1}$ for the fastest hydroxyl groups
\cite{AMBER94:,Luck:73,Krimm:86}. It appears, therefore, that
our $h_{c}$ value corresponds to $\tau\approx 1.1$ in reasonable agreement
with an ideal harmonic model \cite{Mzjcp:97}.
For a frequency of 3800 cm$^{-1}$ the
stability limit of the leapfrog scheme is $h=2.78$~fsec \cite{Hockney:81},
and in low temperature tests the numerical trajectory remains stable right up
to this value. At normal temperature, however, a significant E-drift
appears with $h>2$~fsec, which indicates that some fast anharmonic motions
occur in the system, and the test trajectory could not be completed because
of an explosive growth in temperature. The high E-drift can be caused,
for instance, by collisions between hydrogens in non-polar contacts.

Since the data presented in Fig. \ref{Ffree} are redundant
each case below is characterized by plots (a) and (b) alone. The deviation of
the total energy is sufficient to evaluate $h_{c}$, while $\bar{U}$ always
behaves similarly to plot (c). As for the value of $D\lbrack U\rbrack$ shown
in Fig. \ref{Ffree}~(d), its large and systematic deviation always results
from the drift of the potential energy which is implicitly included in plot
(b).

\begin{figure}
\centerline{\psfig{figure=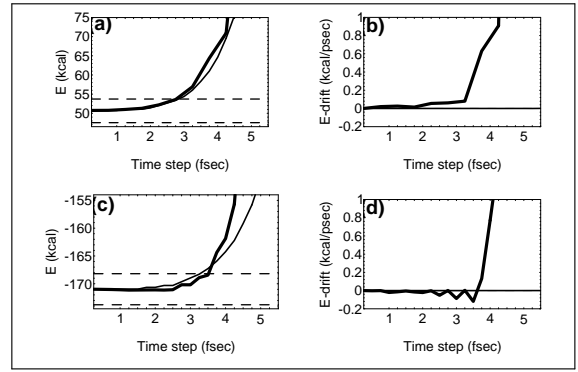,width=7.5cm,angle=0}}
\caption{Protein models with bond length constraints: (a,b) - only bonds to
hydrogen atoms fixed; (c,d) - all bond lengths fixed. The notation is similar
to Figs. \protect\ref{Ffree}~(a) and (b).}\label{Ffxbl}\end{figure}

\subsection*{Proteins with Constrained Bond Lengths}
The two standard modes of constraining bond lengths, that is, constraining only
bonds to hydrogens and constraining all bonds, yield the results shown in
Fig. \ref{Ffxbl}.  We see that they exhibit rather similar behavior with
$h_c$ values around 3~fsec for both low and normal temperature simulations.
In proteins, the fastest bond stretching modes between non-hydrogen
atoms normally occur in carboxyl groups, with frequencies around
1720 cm$^{-1}$~~\cite{Krimm:86}. In our calculations, however,
the fastest mode was found at 1850 cm$^{-1}$ a tryptophan side chain,
which, in a harmonic case, would give a maximum step size of
5.7~fsec and $h_c\approx 3$. We note, therefore, that the plots in Figs.
\ref{Ffxbl} (a,b) agree well with a harmonic approximation and give the
expected $h_c$ value with only bonds to hydrogens constrained. With all
bonds constrained, however, the origin of the limitation is not clear.
The fastest bond angle bending modes are around 1600 cm$^{-1}$~
\cite{Krimm:86}, i.e. not so far, but still they should not impose
limitations in this range of time steps. These limitations are evidently
harmonic, however, because in low temperature tests (see Figs. \ref{Ffxbl}
(a) and (c)) the two models behave almost identically. Note that they both
almost reach the harmonic step size limit, with E-drift close to zero, while
the $h_c$ values are roughly the same as those at normal temperature.

\begin{figure}
\centerline{\psfig{figure=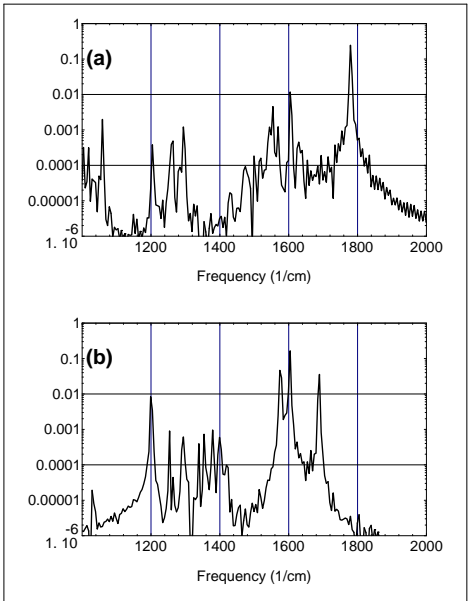,height=8.0cm,angle=0.}}
\caption{Spectral densities of velocity autocorrelation functions for
H--N$_{\epsilon}$--H bond angle of Lys13 for low temperature MD trajectories
calculated with only bonds to hydrogen atoms fixed (a), and with all bond
lengths fixed (b). The spectra had been computed from separate 40~psec runs
with 1~fsec time step and all points stored. The densities are given in
arbitrary scales.}\label{Fspect}\end{figure}

An explanation to this apparently counter-intuitive behavior follows from
the example shown in Fig. \ref{Fspect}. This figure presents spectral
densities of bond bending oscillations of one of the H--N$_{\epsilon}$--H
angles of Lys13 obtained at low temperature with the two different modes of
applying bond length constraints. We see that, with fixed bonds to hydrogen
atoms only, the maximum occurs at 1780~cm$^{-1}$, not very far from the fastest
bond stretching. These two hydrogens are involved
in hydrogen bonds with the peptide
oxygen of Gly9 and a water molecule, respectively. Neither this angle, nor
any of the adjacent bonds or bond angles have independent frequencies above
1700~cm$^{-1}$, nevertheless, the signal at 1780~cm$^{-1}$ is observed in the
spectral densities of many valence and dihedral angles between neighboring
atoms, as well as within the hydrogen bonded Gly9, indicating that
this frequency corresponds to a locally distributed normal mode.

With all bond lengths fixed, this normal mode cannot remain intact, but it
does not disappear, which is clear from Fig. \ref{Fspect}~(b). Compared with
Fig. \ref{Fspect}~(a) there are many fewer signals above 1500~cm$^{-1}$, but
one at 1690~cm$^{-1}$ has appeared. This peak
apparently corresponds to the same fast collective mode as in Fig.
\ref{Fspect}~(a), with bond stretching eliminated, and, as we see, its
frequency is just slightly reduced.
Such behavior is rather characteristic of bond angle vibrations of
hydrogens involved in hydrogen bonding, which explains similar step
size limitations for the two modes of bond length constraint.
A high frequency in this case results from a combination of
several terms in the force field, rather from a single specific one, with
hydrogen bonding as one of the major components. Bond length constraints
affect such modes indirectly, mainly due to redistribution the system inertia.
We note finally that this particular example had been selected because in both
constraint modes there is only one peak above 1650~cm$^{-1}$, therefore,
in the spectrum shown in Fig. \ref{Fspect}~(a), it cannot be attributed to
bond stretching. The largest independent shifts due to hydrogen bonding
are observed in valence angle vibrations of hydroxyl groups and their
frequencies also reach the level of 1700~cm$^{-1}$.

\subsection*{Water Droplet}

\begin{figure}
\centerline{\psfig{figure=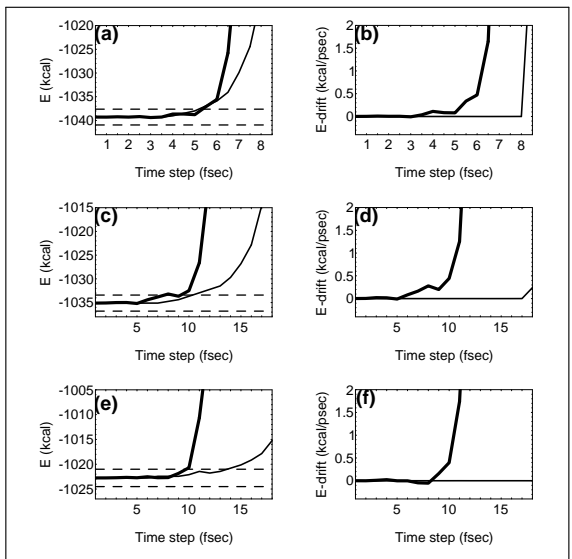,height=7.5cm,angle=0.}}
\caption{Water droplet tests. Results are shown for three different water
models: (a,b) - standard TIP3P water model; (c,d) - added inertia equaled 4;
(e,f) added inertia equaled 15. The notation is as in Figs.
\protect\ref{Ffree}~(a,b).} \label{Fwagb}\end{figure}

The results of tests with a droplet of TIP3P water molecules are shown in
Figs. \ref{Fwagb}~(a,b). They exhibit an $h_c$ value of 5~fsec, although
one should note that beyond 6~fsec, with normal temperature, the E-drift
grows very rapidly. It might be expected that, since in this case no purely
harmonic terms are present in the force filed, the limiting motion should be
anharmonic. The low temperature tests, however, yield exactly the same $h_c$
value and, thus, it appears that the system behaves rather similarly to the
previous ones with a time step limiting frequency of
approximately 1100~cm$^{-1}$. This value is close to the experimental upper
boundary of the band attributed to the rotations of individual water molecules
\cite{Walrafen:72}.

Rotation, or rather libration, of a single water molecule in a net of
hydrogen bonds is certainly the fastest motion here and it can be
specifically slowed down by artificially increasing the moments of inertia
of the water molecule. Such water models present considerable interest for
simulations where structural and thermodynamic properties are targeted,
rather than kinetic ones\cite{Pomes:90}. Figures \ref{Fwagb}~(c,d) demonstrate
results of such tests with an inertia ${\bf I_{ij}=\mu\delta_{ij}}$ added to
oxygen atoms, where $\delta_{ij}$ is Kronecker delta and $\mu=4$ (atom mass
units)$\cdot$\AA$^{2}$. This means that oxygens are no longer considered as
point masses but as spherical rigid bodies of the same mass. With $\mu=4$ the
highest frequency is expected to be reduced approximately by a factor of two,
and it is seen that plots in Figs. \ref{Fwagb}~(c,d) look like those in Figs.
\ref{Fwagb}~(a,b) scaled along the horizontal axis, leading to a
twofold increase in $h_{c}$. Further increase of inertia gives
the effect shown in Fig. \ref{Fwagb}~(e,f) where $\mu$ equals 15.
Instead of an increase proportional
to the square root of the added inertia we obtain $h_{c}$ around 14~fsec and
10~fsec for low and normal temperatures, respectively. In this case,
therefore, we encounter a qualitatively different situation with
significantly anharmonic limitations probably imposed by the translational
motion of water molecules and collisions between them.

By considering the results shown in Fig. \ref{Fwagb} one concludes
that the water model with $\mu =4$ is ``well balanced" for normal temperature
simulations in a sense that both its translational and rotational movements
occur in the same time scale and require similar time steps. It is worth
noting that the step size limits obtained here with the new test
agree well with earlier results. It is known, for instance, that, with a rigid
water model, all liquid structural properties are accurately reproduced up to
a step size of 6 fsec \cite{Fincham:92}. A similar assertion holds for the
``weighted mass" water model up to a step size of 10~fsec\cite{Pomes:90}.

\subsection*{Proteins with Fixed Standard Amino Acid Geometry}

\begin{figure}
\centerline{\psfig{figure=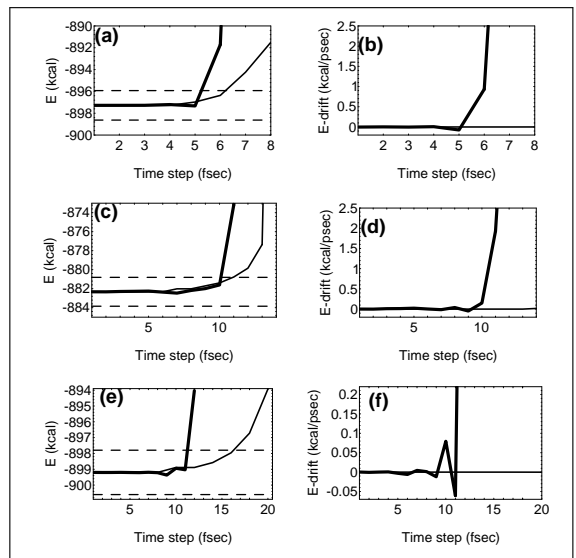,height=7.5cm,angle=0.}}
\caption{Calculations with constraints corresponding to the standard amino
acid geometry. (a,b) - with no modifications of inertia tensors; (c,d) - with
inertia of water molecules and protein hydroxyl groups increased by 4; (e,f)
- with inertia tensors of water molecules and protein hydroxyl and amide
groups increased by 36, and masses of water oxygens equal to 72. The
notation is as in Figs. \protect\ref{Ffree}~(a,b).}\label{Falag}\end{figure}

In these calculations bond lengths and bond angles in the protein were fixed
according to a standard geometry approximation \cite{Go:69}. The results
shown in Figs. \ref{Falag}~(a,b) were obtained with no modifications of inertia
tensors. It is seen that they strongly resemble the water droplet plots in
Figs. \ref{Fwagb}~(a,b). This might mean that the step size limitations are
imposed by a few water clusters around the protein, but calculations with
increased water inertia yield virtually identical results. On the other hand,
simultaneous weighting of water and protein hydroxyl groups yields a
considerable effect as shown in Figs. \ref{Falag}~(c,d). The additional inertia
tensor was same as in the water droplet tests above. Thus, it is evident
that, in the standard geometry representation, libration of hydroxyl groups
is the fastest motion, certainly due to their small inertia. The high
frequency of these librations is due to hydrogen bonding rather than to the
corresponding torsional potential which produces oscillations with at least
three times lower frequencies.

We saw above that a weighted water model with $\mu =4$ gives $h_{c}\approx
10$~fsec, and that this value already does not depend upon the dynamics of
hydrogen motion. It seems that, in general, the 10~fsec level is
characteristic of systems in which the time scale difference between the
movements of hydrogen and non-hydrogen atoms is somehow smoothed, which is
illustrated by Figs. \ref{Falag}~(e,f). In these calculations, in order to
remove all the limitations imposed by water, the masses and inertia tensors
of water oxygens were increased by 56 and 36, respectively, which gives a
fourfold increase of both masses and inertia tensors compared with Figs.
\ref{Fwagb}~(c,d) and, accordingly, a twofold increase in step sizes. The
same inertia was added to protein hydroxyl and amide groups. In the low
temperature plots in Figs. \ref{Falag}~(e,f), it is seen that the harmonic
$h_{c}$ is shifted compared to Fig. \ref{Falag}~(c), which means that, in the
previous model, librations of hydroxyl and amide groups produce the highest
frequencies. The effect, however, takes place at low temperature only, and
the characteristic time step in Fig. \ref{Falag}~(e) is approximately the
same as in Fig. \ref{Falag}~(c). A similar behavior is observed in
calculations with weighted inertia of other hydrogen-only rigid bodies
(results not shown). The situation, therefore, appears very similar to that
in the water droplet tests in Fig. \ref{Fwagb}, but in the present case fast
collisions between non-hydrogen protein atoms present the limiting factor.
A possible way to overcome this limitation is the RESPA
approach \cite{Tuckerman:92} which until now has been used only within the
context of Cartesian MD, but can be adapted for ICMD with the
Hamiltonian equations \cite{Mzjcc:97} since they make possible symplectic
numerical integration.

\section{Concluding remarks}

Although the problem of time step limitations in MD simulations of biopolymers
is long standing, and although the search for effective remedies has been the
subject of many studies, a precise account of the effects that can create
such limitations was missing in the literature. It is necessary to fill
this gap in order that efforts put into the development of new
methods can be effective.

The conventional view of this problem is that the
step size is limited by fast harmonic motions and that these motions are
produced by the stiff components of empirical potentials connected with the
deformations of bonds, bond angles, planar groups etc. It has been shown here
that this description takes into account only a part of step limiting
factors. First, there may be many fast harmonic vibrations in which hydrogen
bonding plays a major role. Second, non-bonded inter-atom interactions impose
ubiquitous anharmonic limitations starting from rather small step sizes. All
such factors form a complicated hierarchy which slightly differs between
different empirical potentials and which is modified when
constraints are imposed.

Our results suggest that the time steps currently used in MD
simulations can be increased considerably with no serious loss in the
accuracy of thermodynamic averages. In unconstrained calculations with
AMBER potentials the estimated limit is about 1.7 fsec. Note, however,
that this value is mainly connected with the frequency of OH-stretching
which is high in AMBER because it close to the upper
limit of experimental frequencies of free hydroxyl groups \cite{Luck:73}.
In the ENCAD potentials \cite{Levitt:97}, for instance, the corresponding force
constant is lowered by a factor of 2, which is why a larger step size of
2 fsec recommended by the authors is safe and could possibly even be
increased further.

Constraining bonds to hydrogen atoms removes the lowest level in the
hierarchy of fast motions and makes possible time steps up to 3 fsec.
Constraints on other bond lengths,
however, are effective only together with constraining bond angles. With
fixed standard amino acid geometry, rotations of hydrogen bonded hydroxyl
groups limit time steps at a 5 fsec level. The next important limitation
occurs around 10 fsec and it is due to collisions between non-hydrogen
protein atoms.

\acknowledgements
I wish to thank R. Lavery for useful comments to the first version of
this paper.


\begin{thebibliography}{10}

\bibitem{SHAKE:}
Ryckaert, J.~P.; Ciccotti, G.; Berendsen, H. J.~C. {\em J. Comput. Phys.} {\bf
  1977}, {\em 23},  327.

\bibitem{vanGunsteren:77}
van Gunsteren, W.~F.; Berendsen, H. J.~C. {\em Mol. Phys.} {\bf 1977}, {\em
  34},  1311.

\bibitem{Ciccotti:86}
Ciccotti, G.; Ryckaert, J.~P. {\em Comput. Phys. Rep.} {\bf 1986}, {\em 4},
  345.

\bibitem{Pinches:78}
Pinches, M. R.~S.; Tildesley, D.~J.; Saville, G. {\em Mol. Phys.} {\bf 1978},
  {\em 35},  639.

\bibitem{Tuckerman:92}
Tuckerman, M.~E.; Berne, B.~J.; Martyna, G.~J. {\em J. Chem. Phys.} {\bf 1992},
  {\em 97},  1990.

\bibitem{Pomes:90}
Pomes, R.; McCammon, J.~A. {\em Chem. Phys. Lett.} {\bf 1990}, {\em 166},  425.

\bibitem{Mao:91}
Mao, B.; Maggiora, G.~M.; Chou, K.~C. {\em Biopolymers} {\bf 1991}, {\em 31},
  1077.

\bibitem{Pear:79}
Pear, M.~R.; Weiner, J.~H. {\em J. Chem. Phys.} {\bf 1979}, {\em 71},  212.

\bibitem{BKS0:}
Mazur, A.~K.; Abagyan, R.~A. {\em J. Biomol. Struct. Dyn.} {\bf 1989}, {\em 6},
   815.

\bibitem{Gibson:90}
Gibson, K.~D.; Scheraga, H.~A. {\em J. Comput. Chem.} {\bf 1990}, {\em 11},
  468.

\bibitem{Jain:93}
Jain, A.; Vaidehi, N.; Rodriguez, G. {\em J. Comput. Phys.} {\bf 1993}, {\em
  106},  258.

\bibitem{Rice:94}
Rice, L.~M.; Br{\"u}nger, A.~T. {\em Proteins: Struct. Funct. Genet.} {\bf
  1994}, {\em 19},  277.

\bibitem{Brooks:88}
Brooks, C.~L., III; Karplus, M.; Pettitt, B.~M. {\em Adv. Chem. Phys.} {\bf
  1988}, {\em 71},  175.

\bibitem{Mathiowetz:94}
Mathiowetz, A.~M.; Jain, A.; Karasawa, N.; Goddard, W.~A., III {\em Proteins:
  Struct. Funct. Genet.} {\bf 1994}, {\em 20},  227.

\bibitem{Mzjcc:97}
Mazur, A.~K. {\em J. Comput. Chem.} {\bf 1997}, {\em 18},  1354.

\bibitem{vanGunsteren:82}
van Gunsteren, W.~F.; Karplus, M. {\em Macromolecules} {\bf 1982}, {\em 15},
  1528.

\bibitem{Dor-jcp:93}
Dorofeyev, V.~E.; Mazur, A.~K. {\em J. Comput. Phys.} {\bf 1993}, {\em 107},
  359.

\bibitem{Mzjcp:97}
Mazur, A.~K. {\em J. Comput. Phys.} {\bf 1997}, {\em 136},  354.

\bibitem{vanGunsteren:90}
van Gunsteren, W.~F.; Berendsen, H. J.~C. {\em Angew. Chem.} {\bf 1990}, {\em
  29},  992.

\bibitem{pgb:}
Gallagher, T.; Alexander, P.; Bryan, P.; Gilliland, G.~L. {\em Biochemistry}
  {\bf 1994}, {\em 33},  4721.

\bibitem{PDB:}
Bernstein, F.~C.; Koetzle, T.~F.; Williams, G. J.~B.; Meyer, E.~F.; Brice,
  M.~D.; Rodgers, J.~R.; Kennard, O.; Shimanouchi, T.; Tasumi, M. {\em J. Mol.
  Biol} {\bf 1977}, {\em 112},  535.

\bibitem{AMBER:}
Pearlman, D.~A.; Case, D.~A.; Caldwell, J.~C.; Ross, W.~S.; Cheatham, T.~E.,
  III; Ferguson, D.~M.; Seibel, G.~L.; Singh, U.~C.; Weiner, P.~K.; Kollman,
  P.~A. {\em AMBER 4.1}; University of California: San Francisco, 1995.

\bibitem{TIP3P:}
Jorgensen, W.~L.; Chandreskhar, J.; Madura, J.~D.; Impey, R.~W.; Klein, M.~L.
  {\em J. Chem. Phys} {\bf 1991}, {\em 79},  926.

\bibitem{AMBER94:}
Cornell, W.~D.; Cieplak, P.; Bayly, C.~I.; Gould, I.~R.; Merz, K.~M.; Ferguson,
  D.~M.; Spellmeyer, D.~C.; Fox, T.; Caldwell, J.~W.; Kollman, P.~A. {\em J.
  Amer. Chem. Soc.} {\bf 1995}, {\em 117},  5179.

\bibitem{Allen:87}
Allen, M.~P.; Tildesley, D.~J. {\em Computer Simulation of Liquids}; Clarendon
  Press: Oxford, 1987.

\bibitem{Haile:92}
Haile, J.~M. {\em Molecular Dynamics Simulations: Elementary Methods};
  Wiley-Interscience: New York, 1992.

\bibitem{Hockney:81}
Hockney, R.~W.; Eastwood, J.~W. {\em Computer Simulation Using Particles};
  McGraw-Hill: New-York, 1981.

\bibitem{Luck:73}
Luck, W. A.~P.  in {\em Water - A Comprehensive Treatise}; Franks, F., Ed.;
  Plenum: New York, 1972; Vol.~2, pp.\ 151--214.

\bibitem{Krimm:86}
Krimm, S.; Bandekar, J. {\em Adv. Prot. Chem.} {\bf 1986}, {\em 38},  181.

\bibitem{Walrafen:72}
Walrafen, G.~E.  in {\em Water - A Comprehensive Treatise}; Franks, F., Ed.;
  Plenum: New York, 1972; Vol.~1, pp.\ 151--214.

\bibitem{Fincham:92}
Fincham, D. {\em Mol. Simul.} {\bf 1992}, {\em 8},  165.

\bibitem{Go:69}
G{\= o}, N.; Scheraga, H.~A. {\em J. Chem. Phys.} {\bf 1969}, {\em 51},  4751.

\bibitem{Levitt:97}
Levitt, M.; Hirshberg, M.; Sharon, R.; Laidig, K.~E.; Daggett, V. {\em J. Phys.
  Chem.} {\bf 1997}, {\em 101B},  5051.

\end{thebibliography}
\end{document}